\documentclass{article}
\usepackage{ismir,amsmath,cite,url}
\usepackage{graphicx}
\usepackage{changepage}
\usepackage{color}
\graphicspath{{./figs/}}

\title{Generating Nontrivial Melodies for Music as a Service}

\threeauthors
{Yifei Teng} {U. of Illinois, Dept. of ECE \\ {\tt teng9@illinois.edu}}
{Anny Zhao} {U. of Illinois, Dept. of ECE \\ {\tt anzhao2@illinois.edu}}
{Camille Goudeseune} {U. of Illinois, Beckman Inst. \\ {\tt cog@illinois.edu}}

\usepackage{algorithm}
\usepackage[noend]{algpseudocode}
\usepackage{caption}
\usepackage{xspace}
\algrenewcommand{\algorithmicrequire}{\textbf{Input:}}
\algrenewcommand{\algorithmicensure}{\textbf{Output:}}
\algrenewcommand{\algorithmicforall}{\textbf{for each}}

\DeclareMathOperator*{\argmin}{argmin} 
\usepackage{mathtools}
\usepackage{multirow}
\usepackage{bigdelim}
\usepackage[table,xcdraw]{xcolor}
\usepackage{neuralnetwork}
\usepackage{tikz}
\usepackage{color, colortbl}
\usepackage{pgfplots}
\usepackage{abc}

\usetikzlibrary{arrows,positioning}
\tikzset{
    >=stealth',
    punkt/.style={
            rectangle,
            rounded corners,
            draw=black, very thick,
            text width=7.5em,
            minimum height=2em,
            text centered},
    nn/.style={
            rectangle,
            rounded corners,
            draw=black, very thick,
            text width=7.5em,
            minimum height=1.5em,
            text centered},
    chordstext/.style={
            text centered,
            text width=3.5em},
    rnn/.style={
            rectangle,
            draw=black, thick,
            text width=7.5em,
            minimum height=1.5em,
            text centered},
    timeconv/.style={
            ellipse,
            draw=black, thick,
            text width=9.5em,
            minimum height=1.5em,
            text centered},
    pil/.style={
            ->,
            thick,
            shorten <=2pt,
            shorten >=2pt,},
    nnarr/.style={
            ->,
            shorten <=1pt,
            shorten >=1pt,}
}

\definecolor{Gray}{gray}{0.9}
\definecolor{DarkerGray}{gray}{0.7}

\begin{document}
\maketitle

\begin{abstract}
We present a hybrid neural network and rule-based system that generates pop music.
Music produced by pure rule-based systems often sounds mechanical.
Music produced by machine learning sounds better,
but still lacks hierarchical temporal structure.
We restore temporal hierarchy by augmenting machine learning with a temporal production grammar,
which generates the music's overall structure and chord progressions.
A compatible melody is then generated by a conditional variational recurrent autoencoder.

The autoencoder is trained with eight-measure segments from
a corpus of 10,000 MIDI files, each of which has had its
melody track and chord progressions identified heuristically.

The autoencoder maps melody into a multi-dimensional feature space,
conditioned by the underlying chord progression.
A melody is then generated by feeding a random sample from that space
to the autoencoder's decoder,
along with the chord progression generated by the grammar.
The autoencoder can make musically plausible variations on
an existing melody, suitable for recurring motifs. It can also reharmonize
a melody to a new chord progression, keeping the rhythm and contour.

The generated music compares favorably with that generated by other academic and commercial software
designed for the music-as-a-service industry.
\end{abstract}

\section{Introduction}\label{sec:introduction}

Computer-generated music has started to expand from its pure artistic and
academic roots into commerce.  Companies such as Jukedeck and Amper
offer so-called music as a service, by analogy with software as a service.
However, their melodies, when present at all, often just arpeggiate the underlying chord.

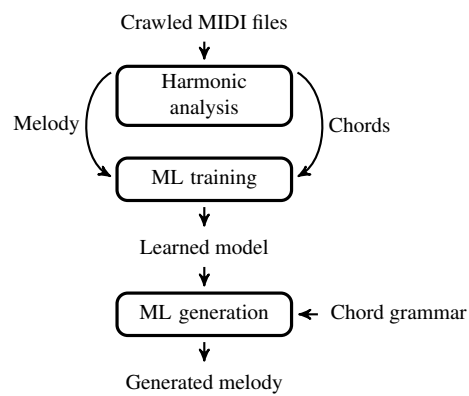
\begin{figure}
\begin{center}
\begin{tikzpicture}[node distance=5mm, auto]
  \begin{scope}[scale=0.8,transform shape]
    \node[] (dummy) {};
    \node[below=of dummy] (dummy2) {};
    \node[punkt, below=of dummy2] (learn) {ML training};
    \node[above=of dummy] (midi) {Crawled MIDI files};
    \node[punkt, below=of dummy2, above=of learn] (anal) {Harmonic analysis};
    \node[below=of learn] (model) {Learned model};
    \path[->] (midi)  edge[pil] (anal);
    \path[->] (anal)  edge[pil, bend right=60] node [left] {Melody} (learn.west);
    \path[->] (anal)  edge[pil, bend left=60] node {Chords} (learn.east);
    \path[->] (learn) edge[pil] (model);

    \node[punkt, below=of model] (gener) {ML generation};
    \node[right=of gener] (tggg) {Chord grammar};
    \node[below=of gener] (mmm)  {Generated melody};
    \path[->] (model) edge[pil] (gener);
    \path[->] (tggg)  edge[pil] (gener);
    \path[->] (gener) edge[pil] (mmm);
  \end{scope}
\end{tikzpicture}
\end{center}
\caption{Machine learning (ML) workflow for generating music from a MIDI corpus.}
 \label{fig:workflow}
\end{figure}

We extend this approach by generating music with both chord progressions
and interesting, nontrivial melodies.
We expand a song structure such as $A A' B A$ into a harmonic plan,
and then add a melody compatible with this structure and harmony.
This compatibility uses a chord-melody relationship found by applying
machine learning to a corpus of MIDI transcriptions of pop music
(\figref{fig:workflow}).

\vspace{1em}

Prior research is discussed in section \ref{sec:related_work}.
Harmonic analysis is detailed in sections \ref{sec:melody_identification} and \ref{sec:chord_detection}.
Hierarchy generation and melody generation are described in section \ref{sec:machine_learning}.

\section{Related Work}\label{sec:related_work}

Recent approaches to machine composition use neural networks (NNs),
hoping to approximate how humans compose.
Chu et al~\cite{1611.03477} generate a melody with a hierarchical NN
that encodes a composition strategy for pop music, and then accompany the
melody with chords and percussion.  However, this music lacks hierarchical
temporal structure.
Boulanger-Lewandowski et al~\cite{1206.6392} investigate hierarchical
temporal dependencies and long-term polyphonic structure.
Inspired by how an opening theme often recurs at a song's end,
they detect patterns with a recurrent temporal restricted Boltzmann machine (RTRBM).
This can represent more complicated temporal distributions of notes.
Similarly, Huang and Wu~\cite{1606.04930} generate structured music
with a 2-layer Long Short Term Memory (LSTM) network.
Although the resulting music often sounds plausible,
it cannot produce clearly repeated melodic themes, just like a Markov
resynthesis of the text of the famous poem ``Jabberwocky'' is unlikely
to replicate the identical opening and closing stanzas of the original.
Despite the LSTM network's theoretical capability of long-term memory, it
fails to generalize to arbitrary time lengths~\cite{1410.5401},
and its generated melodies remain unimaginative.

In these approaches, tonic chords dominate, and melody is little more than arpeggiation.
To avoid this banality, we work in reverse.
We first create structure and chords, and then fit melody to that.
This mimics how classical western Roman-numeral harmony
is taught to beginners:
only after one has the underlying chord sequence,
can one explain the melody in terms of
chord tones, passing tones, appoggiaturas, and so on.


\section{Melody identification}\label{sec:melody_identification}

For pop music, a catchy and memorable melody is crucial.
To generate melodies that sound less robotic than those generated
by other algorithms, we use machine learning.
To create a learning database, we started with a corpus of 10,000 MIDI files~\cite{composing_ai},
from which we extracted useful training data
(melodies that sound vivid or fun).
In particular, the training data was eight-measure excerpts labelled as melody and chords.
We thus had to identify which of a MIDI file's several tracks contained the melody.
To do so, we assigned each track the sum of a rubric score and an entropy score.
Whichever track scored highest was declared to be the melody.
(Ties between high-scoring tracks were broken arbitrarily,
because they were usually due to several tracks having identical notes,
differing only in which instrument played them.)

\subsection{Rubric Score}\label{subsec:rubric}

Our rubric considered attributes such as
instrumentation, note density, and pitch range.

We first considered a track's instrument name (MIDI meta-event {\tt FF} {\tt 04}).
Certain instruments are more common for melody, such as violin or flute.
Others are more likely to be applied as accompaniment or long sustained notes,
such as low brass.
A third category is likely used as unpitched percussion.
The instrument's category then adjusted the rubric's score.

We also considered the track's note density, how often at least one
note is sounding (between corresponding MIDI note-on and note-off events),
as a fraction of the track's full duration.  A track scored higher if
this was between 0.4 and 0.8, a typical value for pop melodies.

Finally we considered pitch range, because we observed that pop melodies often lie between C3 and C5.
The score was higher for a pitch range between C3 and C6,
to exclude bass tracks from consideration.

The values for these attributes were chosen based on manual inspection of 100 files in the corpus.

\subsection{Entropy score}\label{subsec:entropy}

We empirically observed that melody tracks often have a greater variety of pitches than other tracks.
Thus, to quantify how varied, complex, and dynamic a track was,
we calculated each track's entropy
\begin{equation}
  H(X) = - \sum\nolimits_{i=1}^{12} {P(x_i) \log P(x_i)}
\end{equation}
where $x_i$ represents the event that a particular note in the octave is $i$ semitones from the pitch C,
and $P(x_i)$ represents that event's probability.
Higher entropy corresponds to a greater number of distinct pitches.

\subsection{Evaluation}
To measure how well this scoring identified melody tracks,
we manually tagged the melody track of 160 randomly selected MIDI files.
Comparing the scored prediction to this ground truth showed that the error rate was 15\%.

\section{Chord Detection}\label{sec:chord_detection}

To identify the chords in a MIDI file, we considered three aspects of how
pop music differs from genres like classical music.
First, chord inversions (where the lowest note is not the chord's root)
are rare.  When a new chord is presented, it is often in root position:
most pop songs have a clear melody line and bass line~\cite{matti},
and the onset of a new chord is marked with the chord's root in that bass line.
Second, chords may contain extensions (seventh), substitutions (flattened fifth),
doublings, drop voicings (changing which octave a pitch sounds in), and omissions (third or fifth).
Although such modifications complicate the task of functional harmony analysis,
this is not a concern for our application.
Third, new chord onsets are often at the start of a measure; rarely are there
more than two chords per measure.
Combining these observations led us to the following chord detection algorithm.

We first partition the song into segments with constant time signatures.
(these are explicitly stated as MIDI meta messages).  Then each segment
is evenly divided into bins, where we try to match the entire bin to a chord.
Because chords have different durations, we try different bin lengths:
half a measure, one measure, and two measures.
Then for each bin, containing all the notes sounding during that time interval,
we add all these notes to a set
that is matched against a fixed collection of chords,
based on how close the pitches are, with a cost function:


\begin{algorithm}
    \caption*{\textbf{Chord Detection:} \textsc{Cost}} \label{alg:ChordScoring}
    \begin{algorithmic}[1]
        \Function{BestChordInBin}{$Pitches$}
            \State $Root \gets \text{Lowest note starting before first upbeat}$
            \State $Chords \gets \text{All chords, as array of intervals}$\\
            \Return $\argmin_{C \in Chords}\{ \Call{Cost}{Pitches, C, Root} \}$
        \EndFunction
        \Function{Cost}{$Pitches, Chord, Root$}
            \State $PitchCost \gets 0$
            \For {$P \in Pitches$}
                \State $interval \gets \text{No. semitones of }P\text{ from }Root$
                \State $d \gets \min_{voice \in Chord}\{ dist(interval, voice) \}$
                \State $PitchCost \gets PitchCost + d$
            \EndFor
            \State $ChordCost \gets 0$
            \For {$voice \in Chord$}
                \State $d \gets \min_{P \in Pitches}\{ dist(P - Root, voice) \}$
                \State $ChordCost \gets ChordCost + d$
            \EndFor\\
            \Return $PitchCost + ChordCost$
        \EndFunction
    \end{algorithmic}
\end{algorithm}

Each chord's cost is the sum of
the distance of the nearest interval in the chord (from the root)
to each interval in the input pitches,
and the distance of the nearest interval in the input pitches (from its ``root'')
to each interval in the chord, based on some definition of distance.
The cost function then returns the lowest-cost chord.

Defining the distance in terms of mere pitch difference in semitones would be simple,
but performs poorly.
For example, matching the pitch set $[C, E, G]$ to the chord $[C, E\flat, G\flat]$
would yield a cost of two, which is far too low.
Instead, our distance function reflects how compatible intervals are.
The unison is the most compatible, with distance zero;
fourths and fifths are next, with distance one (\tabref{tab:interval_compatibility}).
This conveniently handles omitted-fifth chords,
because the chord's root matches the omitted fifth with a distance of only one.

\begin{table}[]
\begin{center}
\begin{tabular}{|l|l|}
\hline
\rowcolor{DarkerGray}
{Distance in semitones} & {Compatibility distance} \\ \hline
0                       & 0                    \\ \hline
1                       & 6                    \\ \hline
2                       & 2                    \\ \hline
3                       & 2                    \\ \hline
4                       & 2                    \\ \hline
5                       & 1                    \\ \hline
6                       & 4                    \\ \hline
7                       & 1                    \\ \hline
\end{tabular}
\caption{Interval compatibility.}
\label{tab:interval_compatibility}
\end{center}
\end{table}

\begin{figure}
\begin{center}
  \abcinput[options={-Oscores/= -c -s 1.0}]{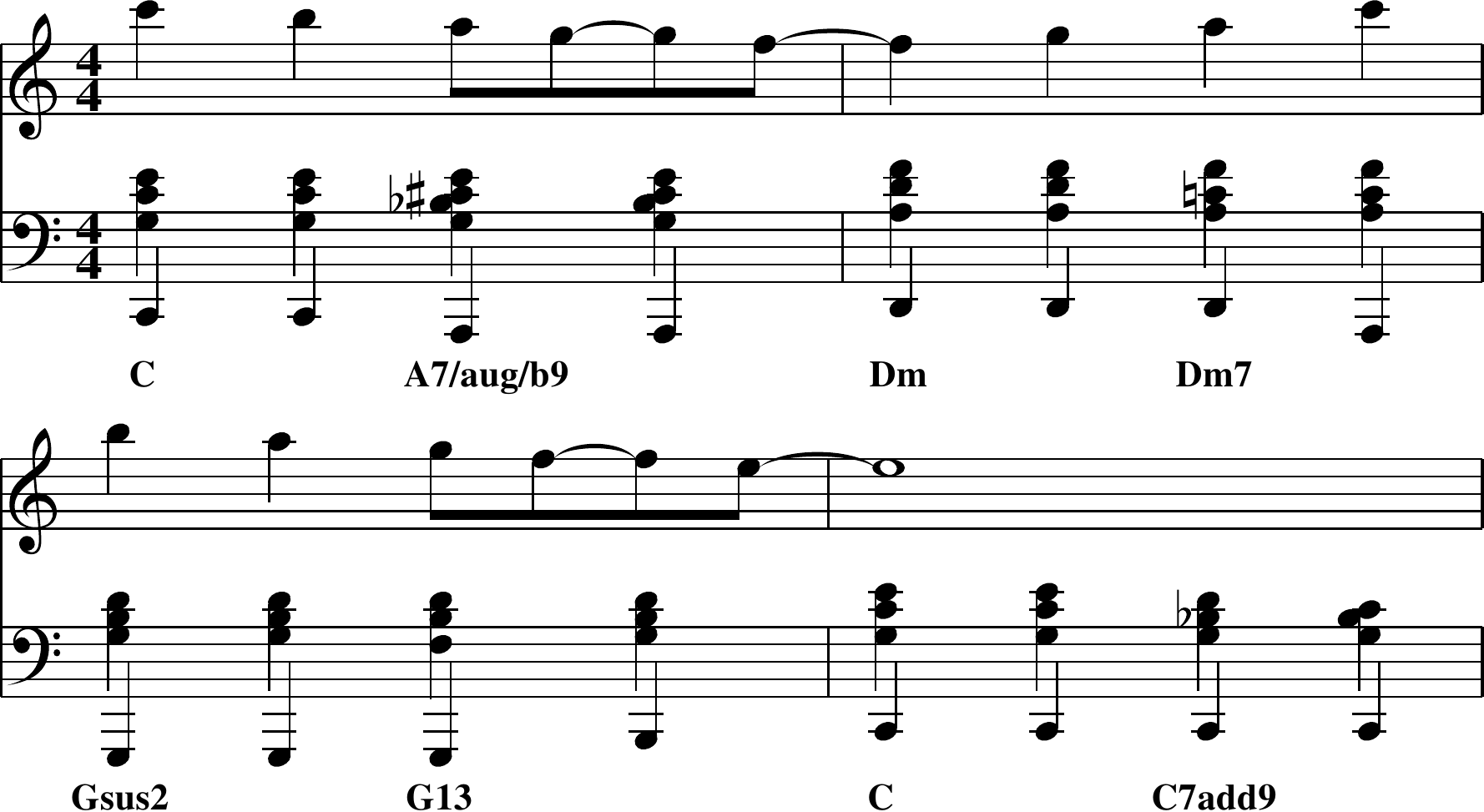}
  \caption{Example of chord detection.}
  \label{fig:fly-me-to-the-moon}
\end{center}
\end{figure}

\figref{fig:fly-me-to-the-moon} demonstrates
chord detection on the song \textit{Fly me to the moon}.
The bin size is half a measure, yielding 8 identified chords.
The $A7/aug/\flat9$ chord resulted from
the accompaniment notes $A, G, B\flat (\text{flat ninth}), C\sharp, E$,
and the melody notes $A, G, F (\text{augmented fifth})$.


\section{Writing Music}\label{sec:machine_learning}

To output pieces with audibly hierarchical structure,
we start with the harmonic structure produced by a temporal generative grammar.
Then an autoencoder recurrent neural network (RNN)
generates a melody compatible with this harmonic scaffold.
The RNN learns to play using the chord's notes,
with occasional surprising non-chord tone decorations
such as passing tones and appoggiaturas.

\subsection{Generating Melody}\label{subsec:cvae_melody}

We first search for a representation of the melody using ML.
This is traditionally done by an autoencoder,
a pair of NNs that maps high-dimensional input data
to and from a lower-dimensional space.
Although this dimensionality reduction can eliminate perfect mappings,
this turns out not to be a problem
because the subspace of ``pleasant'' music within all possible musics is sufficiently small.
Thus, the autoencoder can extract the pleasant content
and map only that into the representation space.

It is tempting to feed a random point from the representation space to the
autoencoder's decoder, and observe how much sense it makes of that point.
However, because one cannot control the shape of the
distribution of melody representations, one cannot guarantee
that a given point from the representation space would be similar to
those seen by the decoder during training.
Thus, the vanilla autoencoder architecture~\cite{Bengio} is not viable as a generative model.
We propose the following improvements for generating melodies:

\begin{enumerate}
    \vspace{-1mm}
    \item \emph{Condition the NN on the chord progression.}
    The chord progression is provided to the NN at every level,
    so when reproducing a melody, the decoder has access to both
    the representation and the chord progression.
    This is useful because a melody has rhythmic information,
    intervallic content, and contour.
    The decoder can ignore the separately provided harmonic information,
    and use only the melody's other aspects.
    This also lets the representation remain constant while altering the chord progression,
    so the NN can adapt a melody to a changed chord progression,
    such as what happens when a key changes from minor to major.

    \vspace{-1mm}
    \item \emph{Add a stochastic layer.}
    Autoencoders which learn a stochastic representation are called variational
    autoencoders, and perform well in generative modelling of images~\cite{1312.6114}.
    The representation is not deterministic. We assume a particular
    (Gaussian) distribution in the representation space, and then train
    the NN to transform this distribution to match the distribution of
    input melodies in their high dimensional space. This ensures that we can
    take a random sampling of the representation space following its associated
    probability distribution, then feed it through the decoder and expect
    a melody similar to the set of musically sensible melodies.

    \vspace{-1mm}
    \item \emph{Use recurrent connections.}
    Pop music has many time-invariant elements, especially at
    time scales below a few beats. A recurrent NN shares
    the same processing infrastructure for note sequences starting at
    different times, and thereby accelerates learning.

    \vspace{-1mm}
    \item \emph{Normalize all other notes relative to the tonic.}
    Pop music is also largely pitch invariant, insofar as a song
    transposed by a few semitones still sounds perceptually similar.
    The NN ignores the song's key and considers the tonic pitch to be abstract,
    as far as pitches in melody and chords are concerned.
    \vspace{-1mm}
\end{enumerate}

\subsubsection{Implementation}\label{subsec:net:impl}

The input melody is quantized to sixteenth notes.  Only sections with an
unchanging duple or quadruple meter are kept. The melody is converted
to a series of one-hot vectors, whose slots represent offsets from
the tonic in the range of $-16$ to $16$ semitones, with one more slot
representing silence.
There is also an attack channel, where
a value of 1 indicates that the note is being rearticulated at the current time step.
The encoding for chords supports up to two chords per measure, and uses a one-hot
vector for scale degrees and separate boolean channels for chord qualities
(\tabref{tab:encoding}). (Note that because this encoding uses just seven Roman-numeral
symbols, it does not try to represent chords outside the current mode.
Before training, we removed from the corpus the few songs that
contained significant occurrences of this.)
We use the basic triad form for each chord identified using techniques from
section \ref{sec:chord_detection}, marking compatible chord qualities.
For example, G$^7$ is encoded by marking a 1 in the $Maj$ and $Pwr$ columns.
(The chord quality encoding could be extended to seventh and ninth chords.)
The table's gray rows are data the network is conditioned on, while the
other rows are input data that the network tries to reproduce.
For an 8-measure example, the input and output vector size is
$35\times8\times16=4480$, and the conditional vector
size is $8\times16+5\times16+8=216$.

\begin{table}[]
\begin{adjustwidth}{-0.7cm}{}
\begin{center}
\begin{tabular}{l|lllll|l}
\cline{2-6}
\ldelim\{{5}{7mm}[16x] & -16                  & \multicolumn{1}{c}{\ldots} & \multicolumn{1}{l|}{16}     & \multicolumn{1}{l|}{Silent}  & Attack                      & \rdelim\}{2}{2.5mm}[x8] \\
 & \multicolumn{1}{c}{\vdots}                 & \multicolumn{1}{c}{\vdots} & \multicolumn{1}{c|}{\vdots} & \multicolumn{1}{c|}{\vdots}  & \multicolumn{1}{c|}{\vdots} \\ \cline{2-6}
 & \cellcolor{Gray} I                         & \multicolumn{1}{c}{\cellcolor{Gray}\ldots} & \multicolumn{1}{l|}{\cellcolor{Gray}VI}     & \multicolumn{1}{l|}{\cellcolor{Gray}VII}     & \cellcolor{Gray}Silent                      \\ \cline{2-6}
 & \multicolumn{1}{l|}{\cellcolor{Gray}Pwr}   & \multicolumn{1}{l|}{\cellcolor{Gray}Maj}   & \multicolumn{1}{l|}{\cellcolor{Gray}Min}    & \multicolumn{1}{l|}{\cellcolor{Gray}Dim}     & \cellcolor{Gray}Aug                         \\ \cline{2-6}
 & \multicolumn{1}{c}{\cellcolor{Gray}\vdots} & \multicolumn{1}{c}{\cellcolor{Gray}\vdots} & \multicolumn{1}{c}{\cellcolor{Gray}\vdots}  & \multicolumn{1}{c}{\cellcolor{Gray}\vdots}   & \multicolumn{1}{c|}{\cellcolor{Gray}\vdots} \\ \cline{2-6}
 & \multicolumn{1}{l|}{\cellcolor{Gray}Major} & \cellcolor{Gray}Dorian                     & \multicolumn{1}{l}{\cellcolor{Gray}\ldots}  & \multicolumn{1}{l|}{\cellcolor{Gray}Locrian}   & \cellcolor{Gray}Jazz Minor                     \\ \cline{2-6}
\end{tabular}
\caption{An encoding of 8 measures (see section \ref{subsec:net:impl}).}
\label{tab:encoding}
\end{center}
\end{adjustwidth}
\end{table}

\begin{figure}
\begin{center}
\begin{tikzpicture}[node distance=4mm, auto]
    \begin{scope}[scale=0.7,transform shape]
      \node[] (dummy) {};
      \node[above=of dummy] (input) {Input vector};

      \node[rnn, below=of dummy] (rec_rnn_1) {600 hidden unit};
      \node[rnn, below=of rec_rnn_1] (rec_rnn_2) {600 hidden unit};
      \node[rnn, below=of rec_rnn_2] (rec_rnn_3) {600 hidden unit};

      \node[rnn, below=of rec_rnn_3] (rec_rnn_4) {600 hidden unit};
      \node[rnn, below=of rec_rnn_4] (rec_rnn_5) {600 hidden unit};
      \node[rnn, below=of rec_rnn_5] (rec_rnn_6) {600 hidden unit};

      \node[rnn, below=of rec_rnn_6] (rec_rnn_7) {600 hidden unit};
      \node[rnn, below=of rec_rnn_7] (rec_rnn_8) {600 hidden unit};
      \node[rnn, below=of rec_rnn_8] (rec_rnn_9) {600 hidden unit};

      \node[rnn, below=of rec_rnn_9] (rec_rnn_a) {600 hidden unit};
      \node[rnn, below=of rec_rnn_a] (rec_rnn_b) {600 hidden unit};
      \node[rnn, below=of rec_rnn_b] (rec_rnn_c) {600 hidden unit};

      \node[right=1cm of rec_rnn_2] (rec_rnn_2_dummy) {};
      \node[right=of rec_rnn_4] (rec_rnn_4_dummy) {};
      \node[right=of rec_rnn_6] (rec_rnn_6_dummy) {};
      \node[right=of rec_rnn_8] (rec_rnn_8_dummy) {};
      \node[right=1cm of rec_rnn_a] (rec_rnn_a_dummy) {};
      \node[chordstext, right=2cm of rec_rnn_6] (rec_chords) {chords and mode};
      \path[->] (rec_chords) edge[nnarr] (rec_rnn_2_dummy);
      \path[->] (rec_chords) edge[nnarr] (rec_rnn_4_dummy);
      \path[->] (rec_chords) edge[nnarr] (rec_rnn_6_dummy);
      \path[->] (rec_chords) edge[nnarr] (rec_rnn_8_dummy);
      \path[->] (rec_chords) edge[nnarr] (rec_rnn_a_dummy);

      \node[timeconv, below=of rec_rnn_c] (rec_fc) {8x300 to 1x1200};

      \node[nn, below left=0.5cm and -1cm of rec_fc] (rec_mean) {800 FC};
      \node[nn, below right=0.5cm and -1cm of rec_fc] (rec_std) {800 FC};

      \node[below=3cm of rec_rnn_c] (latent) {Latent distribution};

      \node[nn, below=4.5cm of rec_rnn_c] (gen_fw) {600 FC};

      \node[rnn, below=1cm of gen_fw] (gen_rnn_1) {600 hidden unit};
      \node[rnn, below=of gen_rnn_1] (gen_rnn_2) {600 hidden unit};
      \node[rnn, below=of gen_rnn_2] (gen_rnn_3) {600 hidden unit};

      \node[rnn, below=of gen_rnn_3] (gen_rnn_4) {600 hidden unit};
      \node[rnn, below=of gen_rnn_4] (gen_rnn_5) {600 hidden unit};
      \node[rnn, below=of gen_rnn_5] (gen_rnn_6) {600 hidden unit};

      \node[rnn, below=of gen_rnn_6] (gen_rnn_7) {600 hidden unit};
      \node[rnn, below=of gen_rnn_7] (gen_rnn_8) {600 hidden unit};
      \node[rnn, below=of gen_rnn_8] (gen_rnn_9) {600 hidden unit};

      \node[rnn, below=of gen_rnn_9] (gen_rnn_a) {600 hidden unit};
      \node[rnn, below=of gen_rnn_a] (gen_rnn_b) {600 hidden unit};
      \node[rnn, below=of gen_rnn_b] (gen_rnn_c) {600 hidden unit};

      \node[right=1cm of gen_rnn_2] (gen_rnn_2_dummy) {};
      \node[right=of gen_rnn_4] (gen_rnn_4_dummy) {};
      \node[right=of gen_rnn_6] (gen_rnn_6_dummy) {};
      \node[right=of gen_rnn_8] (gen_rnn_8_dummy) {};
      \node[right=1cm of gen_rnn_a] (gen_rnn_a_dummy) {};
      \node[chordstext, right=2cm of gen_rnn_6] (gen_chords) {chords and mode};
      \path[->] (gen_chords) edge[nnarr] (gen_rnn_2_dummy);
      \path[->] (gen_chords) edge[nnarr] (gen_rnn_4_dummy);
      \path[->] (gen_chords) edge[nnarr] (gen_rnn_6_dummy);
      \path[->] (gen_chords) edge[nnarr] (gen_rnn_8_dummy);
      \path[->] (gen_chords) edge[nnarr] (gen_rnn_a_dummy);

      \node[timeconv, below=of gen_rnn_c] (gen_fc) {128x300 to 128x35};

      \node[below=of gen_fc] (output) {Output vector};

      \path[->] (input)      edge[nnarr] (rec_rnn_1);
      \path[->] (input)      edge[nnarr, bend right=50] (rec_rnn_4.west);
      \path[->] (input)      edge[nnarr, bend right=60] (rec_rnn_7.west);
      \path[->] (input)      edge[nnarr, bend right=70] (rec_rnn_a.west);

      \path[->] (rec_rnn_1)  edge[nnarr] node {600} (rec_rnn_2);
      \path[->] (rec_rnn_2)  edge[nnarr] node {600} (rec_rnn_3);
      \path[->] (rec_rnn_3)  edge[nnarr] node {300} (rec_rnn_4);
      \path[->] (rec_rnn_4)  edge[nnarr] node {600} (rec_rnn_5);
      \path[->] (rec_rnn_5)  edge[nnarr] node {600} (rec_rnn_6);
      \path[->] (rec_rnn_6)  edge[nnarr] node {300} (rec_rnn_7);
      \path[->] (rec_rnn_7)  edge[nnarr] node {600} (rec_rnn_8);
      \path[->] (rec_rnn_8)  edge[nnarr] node {600} (rec_rnn_9);
      \path[->] (rec_rnn_9)  edge[nnarr] node {300} (rec_rnn_a);
      \path[->] (rec_rnn_a)  edge[nnarr] node {600} (rec_rnn_b);
      \path[->] (rec_rnn_b)  edge[nnarr] node {600} (rec_rnn_c);
      \path[->] (rec_rnn_c)  edge[nnarr] node {2400} (rec_fc);
      \path[->] (rec_fc)     edge[nnarr] node [left=0.3cm] {800} (rec_mean);
      \path[->] (rec_fc)     edge[nnarr] node [right=0.3cm] {800} (rec_std);
      \path[->] (rec_mean)   edge[nnarr] node [left] {Mean} (latent);
      \path[->] (rec_std)    edge[nnarr] node {Standard Deviation} (latent);

      \path[->] (latent)     edge[nnarr] node {Sampling} (gen_fw);

      \path[->] (gen_fw)     edge[nnarr] (gen_rnn_1);
      \path[->] (gen_fw)     edge[nnarr, bend right=50] (gen_rnn_4.west);
      \path[->] (gen_fw)     edge[nnarr, bend right=60] (gen_rnn_7.west);
      \path[->] (gen_fw)     edge[nnarr, bend right=70] (gen_rnn_a.west);

      \path[->] (gen_rnn_1)  edge[nnarr] node {600} (gen_rnn_2);
      \path[->] (gen_rnn_2)  edge[nnarr] node {600} (gen_rnn_3);
      \path[->] (gen_rnn_3)  edge[nnarr] node {300} (gen_rnn_4);
      \path[->] (gen_rnn_4)  edge[nnarr] node {600} (gen_rnn_5);
      \path[->] (gen_rnn_5)  edge[nnarr] node {600} (gen_rnn_6);
      \path[->] (gen_rnn_6)  edge[nnarr] node {300} (gen_rnn_7);
      \path[->] (gen_rnn_7)  edge[nnarr] node {600} (gen_rnn_8);
      \path[->] (gen_rnn_8)  edge[nnarr] node {600} (gen_rnn_9);
      \path[->] (gen_rnn_9)  edge[nnarr] node {300} (gen_rnn_a);
      \path[->] (gen_rnn_a)  edge[nnarr] node {600} (gen_rnn_b);
      \path[->] (gen_rnn_b)  edge[nnarr] node {600} (gen_rnn_c);

      \path[->] (gen_rnn_c)  edge[nnarr] node {300} (gen_fc);
      \path[->] (gen_fc)  edge[nnarr] (output);

    \end{scope}
\end{tikzpicture}
\end{center}
\caption{Network architecture. Rectangles are bidirectional recurrent
neural network cells. Ellipses are strided time-convolution cells. Rounded
rectangles are fully connected (FC) layers. Numbered arrows indicate a connection's dimension.}
\label{fig:network}
\end{figure}
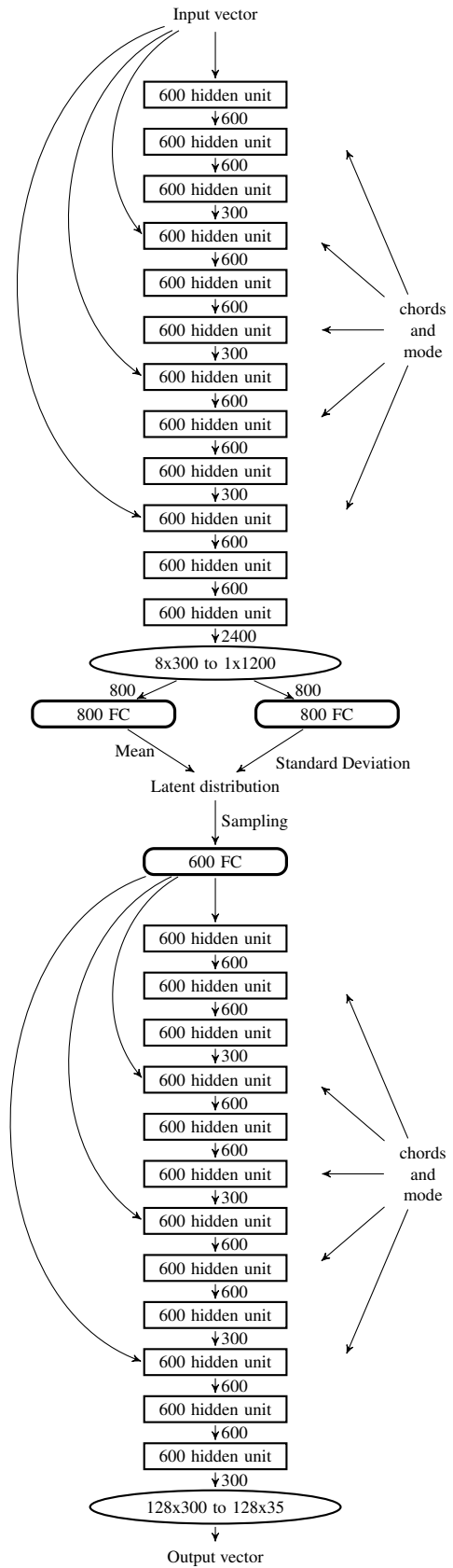

The network has 24 recurrent layers, 12 each for the encoder and decoder (\figref{fig:network}).
Drawing on ideas of deep residual learning from
computer vision~\cite{1512.03385}, we make additional connections from the input
to every third hidden layer. To improve learning, the network accesses both the original melody
and the transformed results from previous layers during processing.
The conditional part (chords and mode) is also provided to
the network at every recurrent layer, as extra incoming connections.

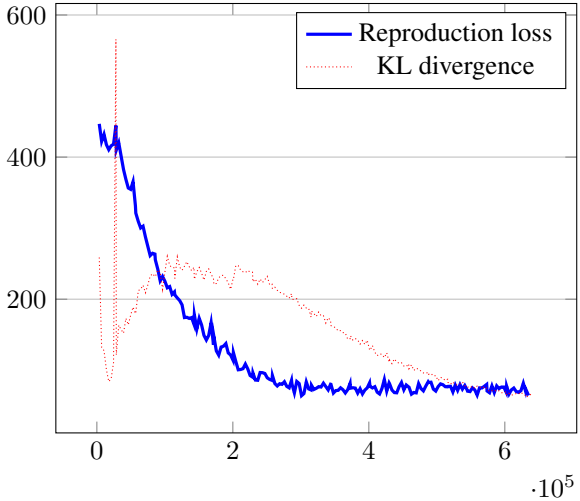
\begin{figure}
    \begin{tikzpicture}
        \begin{axis}[
            mark options={scale=0},
            ymajorgrids=true,
            enlarge y limits,
            try min ticks=5
        ]
            \addplot+[very thick] table [x=Step, y=Value, col sep=comma] {data/reconstr_loss.csv};
            \addlegendentry{Reproduction loss};
            \addplot+[thin, densely dotted] table [x=Step, y=Value, col sep=comma] {data/latent_loss.csv};
            \addlegendentry{KL divergence};
        \end{axis}
    \end{tikzpicture}
    \caption{Training error and Kullback-Leibler divergence of the NN. The horizontal axis indicates how many
    training segments have elapsed ($\times 10^5$).  Initial outliers have been removed.}
    \label{fig:train_error}
\end{figure}

The network is implemented in Tensorflow, a machine learning library for
rapid prototyping and production training~\cite{1603.04467}. It was trained for four days on an
Nvidia Tesla K80 GPU.
We used Gated Recurrent Units~\cite{1406.1078} to build the bidirectional recurrent layer
and Exponential Linear Units~\cite{1511.07289} as activation functions.
These significantly accelerate training
while simplifying the network~\cite{1412.3555,1511.07289}.
\figref{fig:train_error} shows the training error (the sum of model reproduction errors)
and the difference of the latent distribution from a unit Gaussian distribution,
as measured by Kullback-Leibler divergence~\cite{kullback1951}.
The network's input data (available at \texttt{https://goo.gl/VezNNA})
is a set of MIDI songs from various online sources.
Our harmonic analysis converted this to
$1.9\times10^6$ measures of melodies and corresponding chords.
We implemented KL warm up, because that is crucial to learning for
a variational autoencoder~\cite{1602.02282}. But instead of linearly scaling
the KL term for this, we found that a sigmoid reduced the network's reproduction loss.

\subsection{Generating Hierarchy and Chords}\label{subsec:tggg_chords}
Hierarchy and chords are generated simultaneously,
using a temporal generative grammar~\cite{tggg},
modified to suit the harmonies of pop music,
and extended to enable repeated motifs with variations.
The original temporal generative grammar
has notions of sharing by binding a section to a symbol.
For example, the rule
\begin{equation}
\text{let}\ x = \text{I in I } M5(x) \text{ I } M5(x) \text{ I},
\end{equation}
where $M5$ indicates modulating to the $5^\text{th}$ degree,
would expand to five sections,
with the second and fourth identical because $x$ is reused.
We extend this by having symbols $x$ carry along a number: $x_1, x_2, ...$.
Different subscripts of the same symbol still expand to the same chord progression,
but denote slightly different latent representations when
generating corresponding melodies for those sections. The latent representations
corresponding to $x_{i>1}$ are derived from that of $x_1$ by adding random
Gaussian perturbations. This yields variations on the original melody.

\subsection{Training Examples in the Representation Space}\label{subsec:plot_rep}

\begin{figure}
\begin{center}
\includegraphics[scale=0.4]{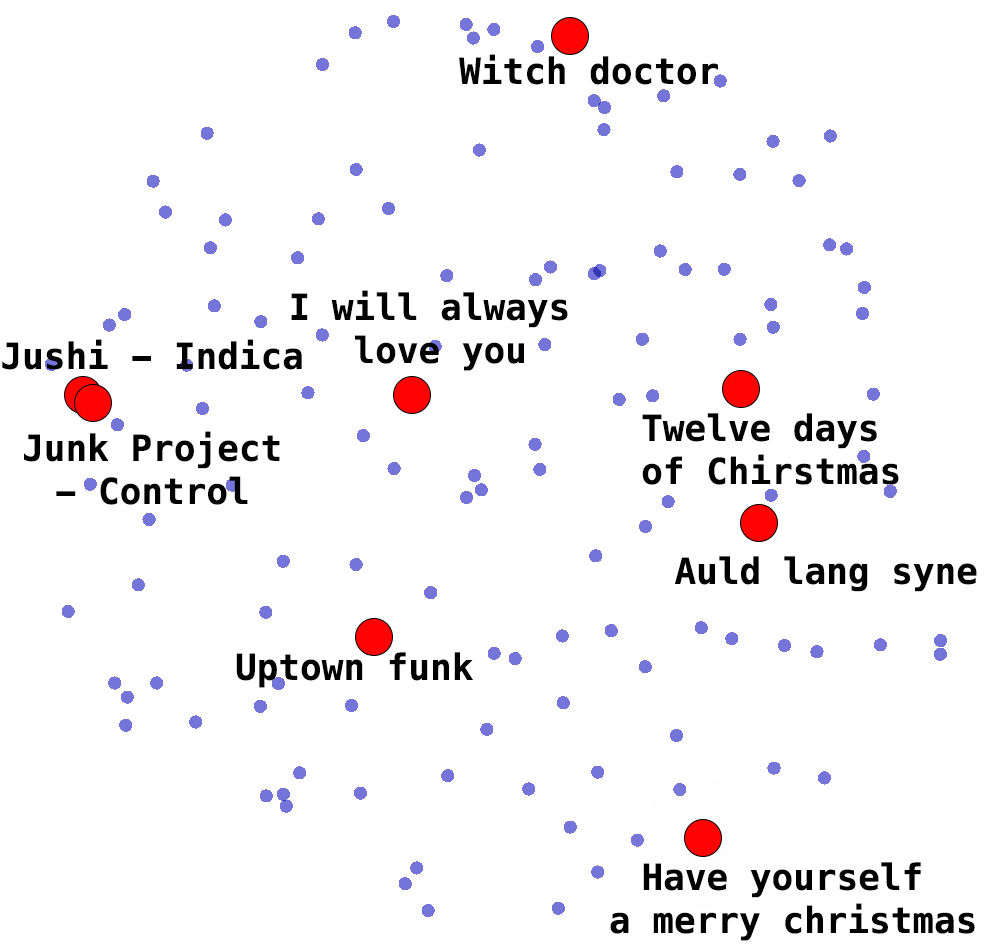}
\caption{Example melodies in a t-SNE plot of the representation space.}
\label{fig:tsne}
\end{center}
\end{figure}

We randomly chose 130 songs from the training set, fed them through the network,
and performed t-SNE analysis on the resulting 130 locations in the representation space.
Although a melody maps to a distribution in the representation space,
\figref{fig:tsne} plots only each distribution's mean, for visual clarity.
This t-SNE analysis effectively reduces the 800-dimensional representation space
into a low-dimensional human-readable format~\cite{ictdbid2777}.
(A larger interactive visualization of 1,680 songs is at \texttt{https://composing.ai/tsne}.)

\begin{figure}
\begin{center}
\vspace{2mm}
\abcinput[options={-Oscores/= -c -s 1.2}]{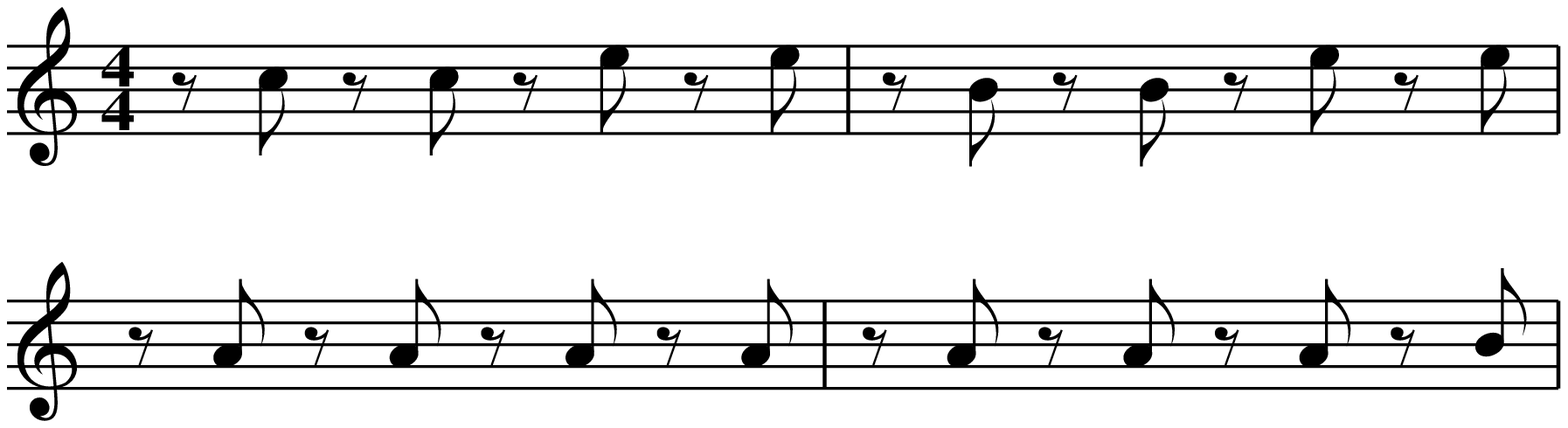}
\vspace{5mm}
\abcinput[options={-Oscores/= -c -s 1.2}]{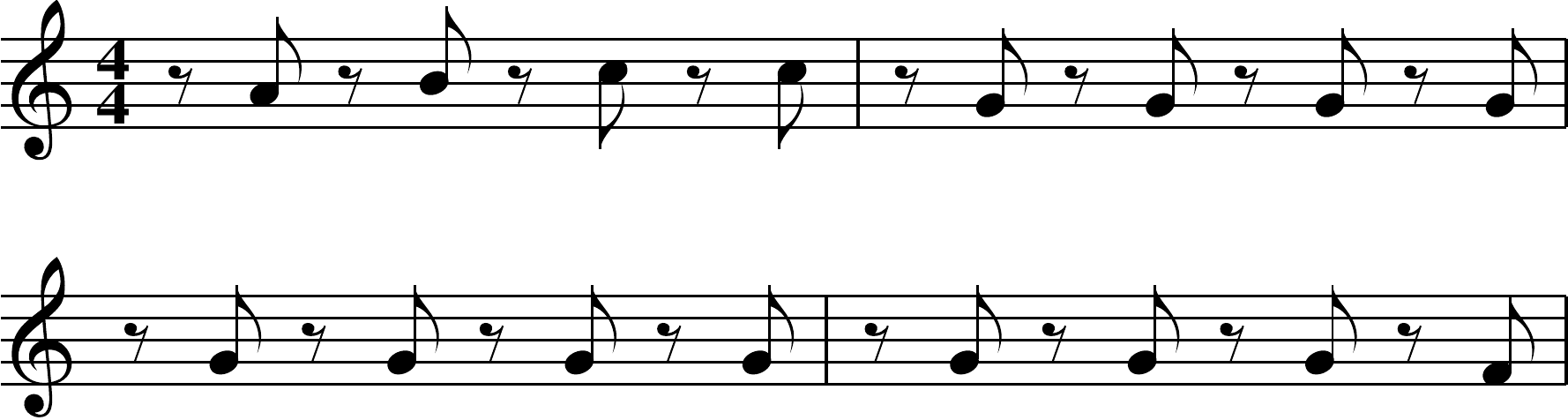}
\vspace{-2mm}
\caption{Four-bar excerpts from the songs \textit{Indica} (top) and \textit{Control} (bottom).}
  \label{fig:indigo_control}
\end{center}
\end{figure}

Two songs that are both in the techno genre,
\textit{Indica} by Jushi and \textit{Control} by Junk Project,
are indeed very near in the t-SNE plot, almost overlapping.
Excerpts from them show that both have a staccato rhythm with notes landing on the upbeat,
and have similar contours (\figref{fig:indigo_control}).

\subsection{Reharmonizing Melody}\label{subsec:change_mood}
\begin{figure}
\begin{center}
\abcinput[options={-Oscores/= -c -s 1.2}]{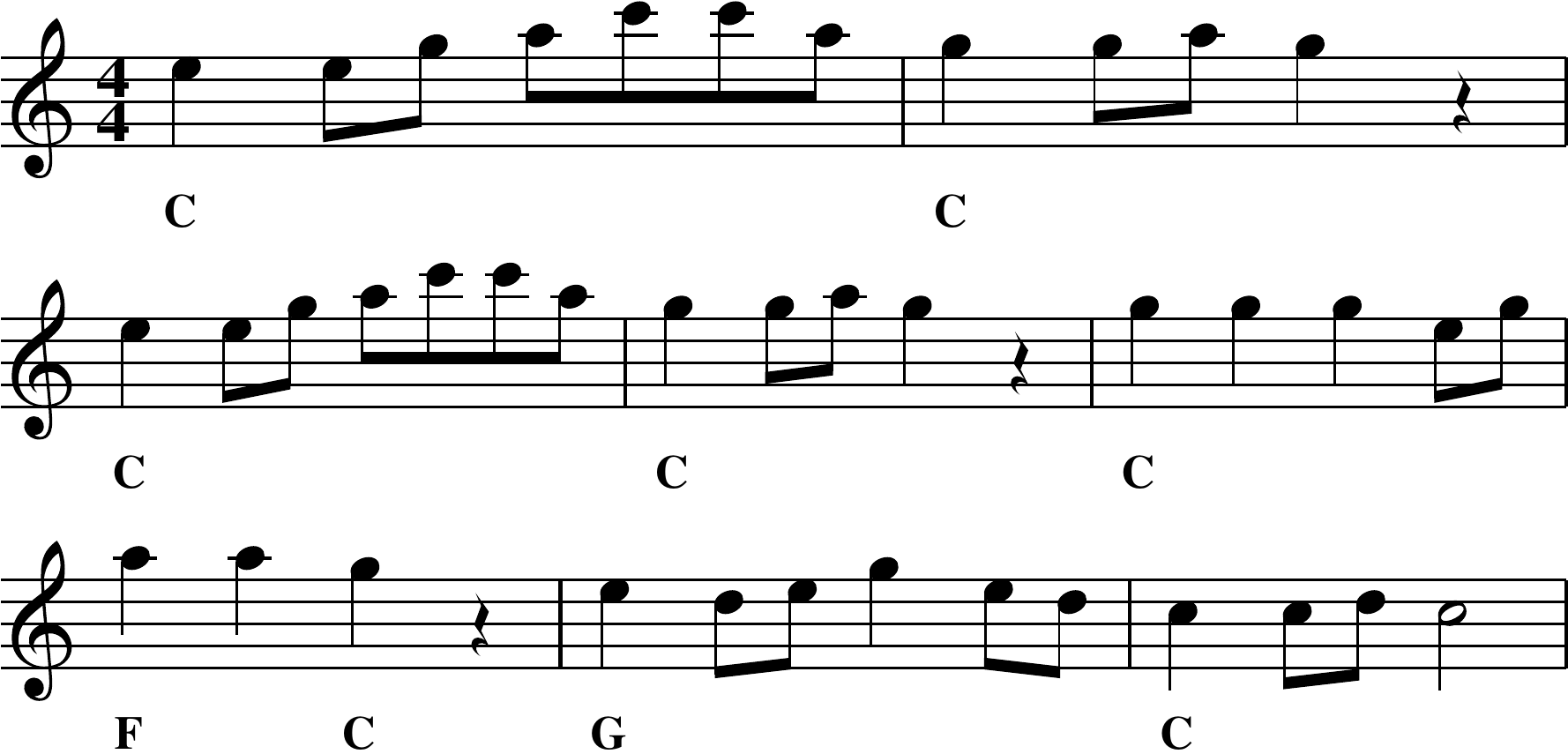}
\vspace{5mm}
\abcinput[options={-Oscores/= -c -s 1.2}]{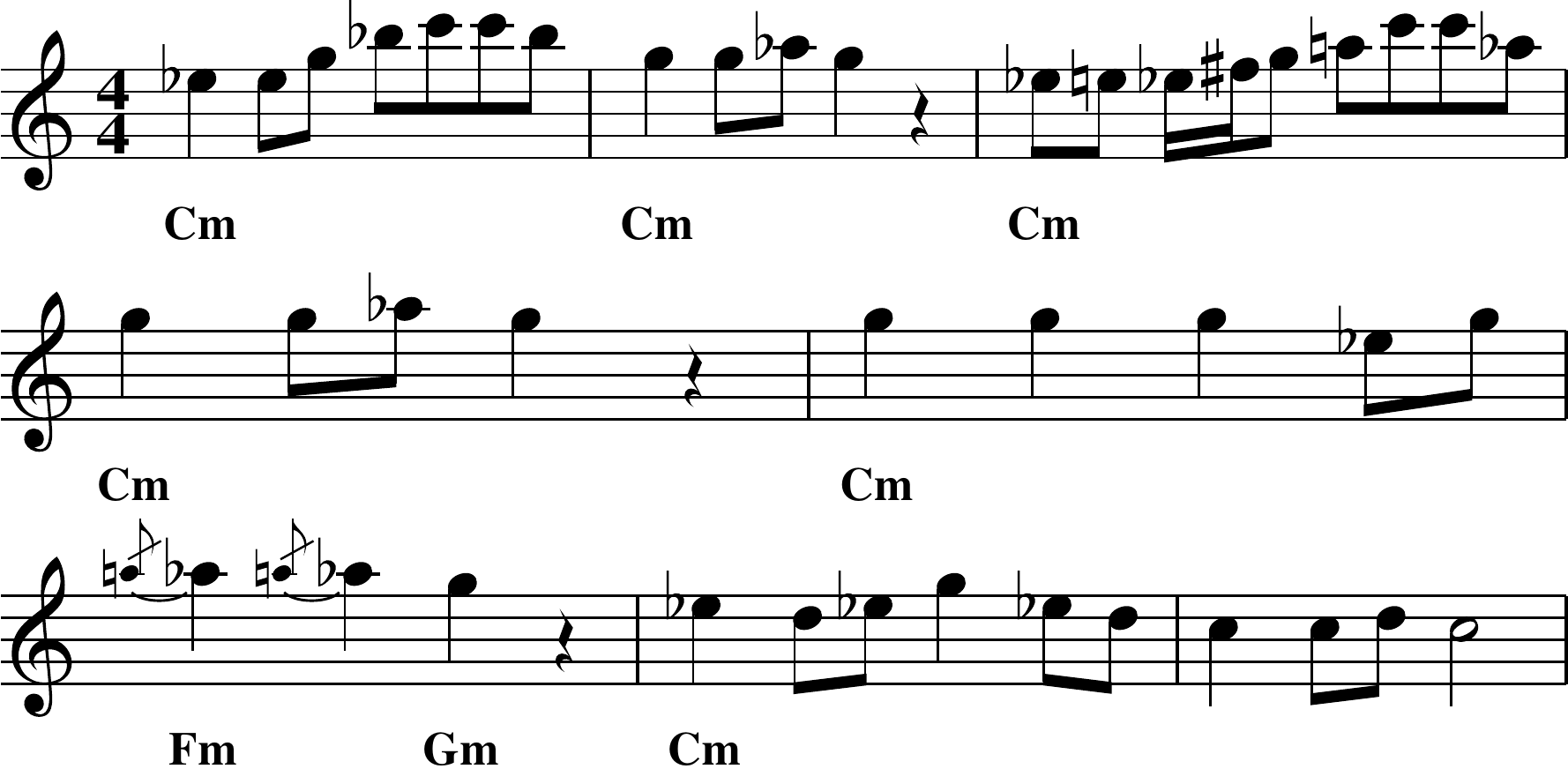}
\vspace{-2mm}
\caption{The song \textit{Jasmine Flower} with original chords (top), and adapted to a new chord progression (bottom).}
\label{fig:jasmine}
\end{center}
\end{figure}

We hypothesized that, when building the neural network architecture,
providing the chord progression to both the encoder and the decoder would
not preserve that information in the representation space,
thus saving space for rhythmic nuances and contour.
To test this hypothesis,
we gave the network songs disjoint from the training set and collected
their representations. We then fed these representations along with
a new chord progression to the network.
We hoped that it would respond by generating a melody that was harmonically
compatible with the new chord progression, while still resembling the original melody.
We demonstrate this with the Chinese folk song \textit{Jasmine Flower},
in a genre unfamiliar to the NN (\figref{fig:jasmine}).
Note that we supplied the chords in \figref{fig:jasmine} (bottom),
for which the NN filled in the melody.
The network flattened the E, A, and B, by observing that the chord
progression looked minor. This is typically how a human would perform the
reharmonization, demonstrating the network's comprehension of how melody and harmony interact.

Although the NN struggled to reproduce the melody,
it provided interesting modifications. The grace notes in measure 6 could be due to
similar ones in the training set, or due to
vacillation between the A$\natural$ from the representation
and the A$\flat$ from the chord conditioning.

\subsection{Examples of generated Melodies}\label{subsec:sample}
Because an entire multi-section composition cannot fit here,
we merely show excerpts from two shorter examples.

\begin{figure}
\begin{center}
  \abcinput[options={-Ogenerated_scores/= -c -s 1.2}]{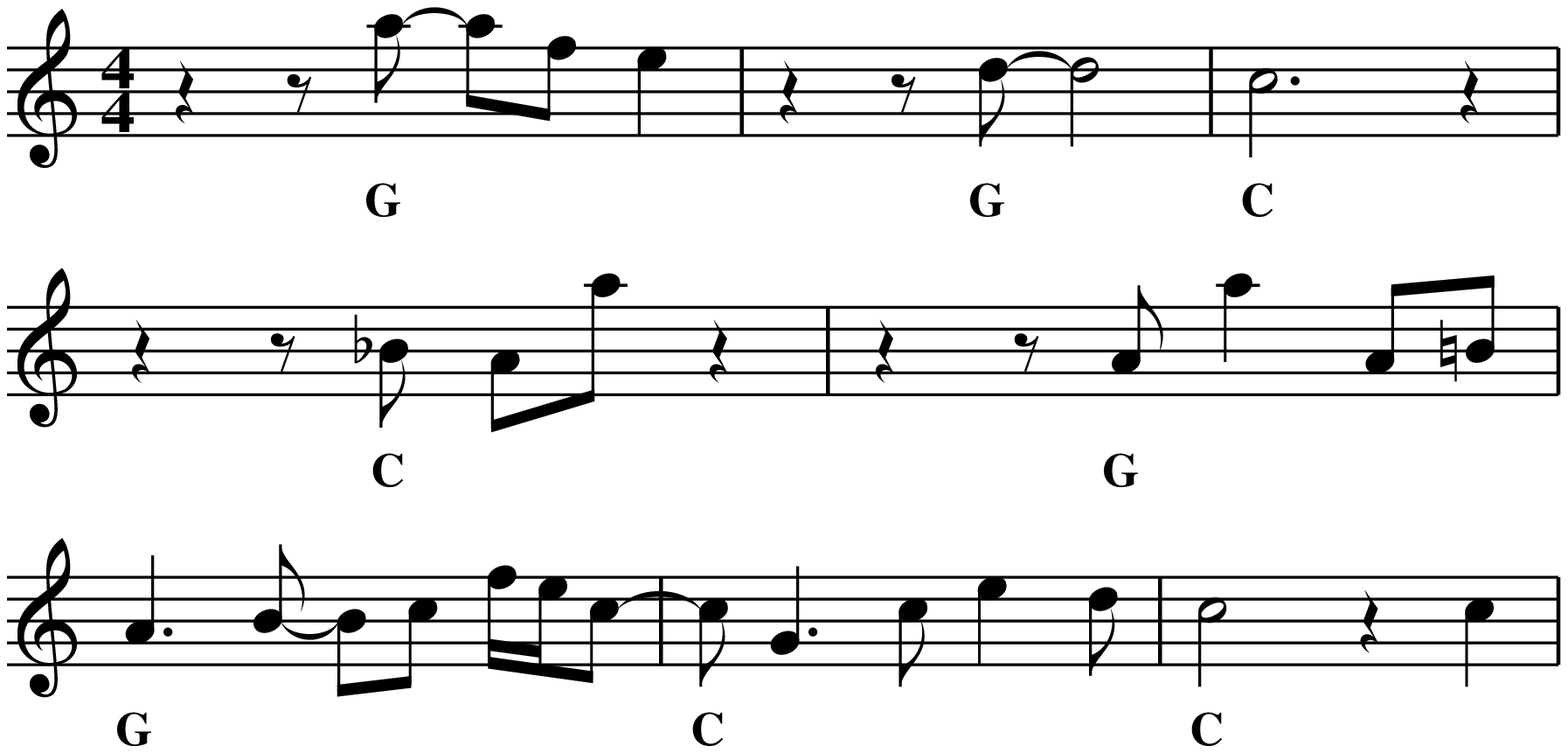}
  \caption{Generated melody for a grammar-generated chord progression.}
  \label{fig:sample33-2}
\end{center}
\end{figure}

\begin{figure}
\begin{center}
  \abcinput[options={-Ogenerated_scores/= -c -s 1.2}]{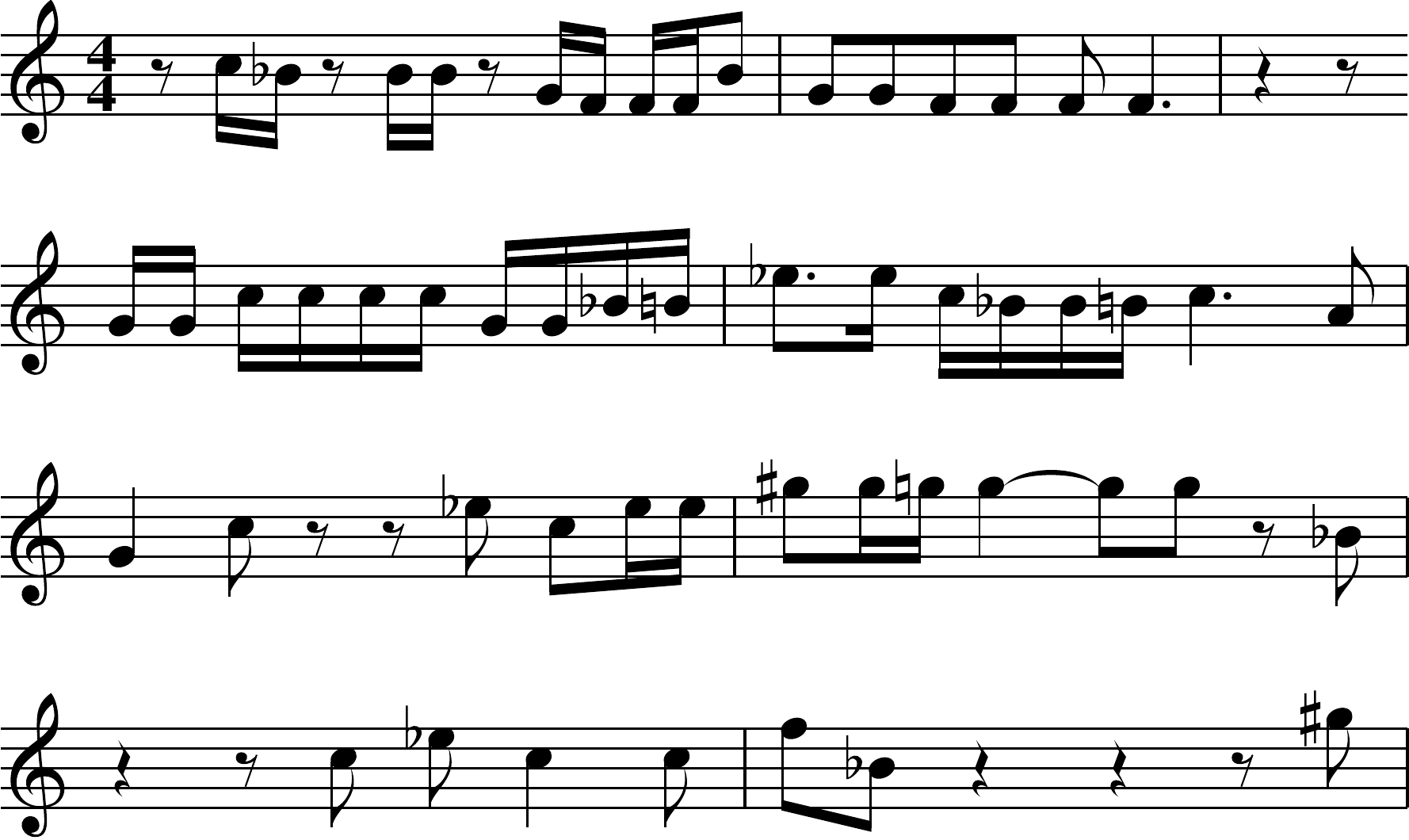}
  \caption{Generated melody for an extended C minor chord.}
  \label{fig:sample33-16}
\end{center}
\end{figure}

\figref{fig:sample33-2} and \figref{fig:sample33-16} demonstrate melodies
generated from points in the representation space that are not near any particular
previously known melody.
Structure is evident in \figref{fig:sample33-2}:
measures 1--3 present a short phrase, and measure 4 leads to the next four measures,
which recapitulate the first three measures with elaborate variation.
\figref{fig:sample33-16} shows an energetic melody where the grammar only produced
C minor chords.  Although the final two measures wander off,
the first six have consistent style and natural contour.

\section{Conclusion and future work}\label{sec:results}

We have combined generative grammars for structure and harmony with
a NN, trained on a large corpus, to emit melodies compatible with
a given chord progression.
This system generates compositions in a pop music style
whose melody, harmony, motivic development,
and hierarchical structure all fit the genre.

This system is currently limited by assuming that the input data's chords
are in root position.  More sophisticated chord detection would still let
it exploit the relative harmonic rigidity of popular music.
Also, by investigating the representation found by the NN,
meaning could be assigned to some of its 800 dimensions, such as intensity, consonance, and contour.
This would let us boost or attenuate a given melody along those dimensions.

\textbf{Acknowledgements.} The authors are grateful to Mark Hasegawa-Johnson for
overall guidance, and to the anonymous reviewers for insights into
both philosophical issues and minute details.

\bibliography{ISMIR}
\end{document}